\documentclass[aps,prl,twocolumn,showpacs,preprintnumbers,amsmath,amssymb,10pt]{revtex4-1} %twocolumn,
\usepackage{amsmath,graphicx,subfig,xcolor,verbatim}
\usepackage{diagbox}

\newcommand{\be}{\begin{equation}}
\newcommand{\ee}{\end{equation}}
\newcommand{\bea}{\begin{eqnarray}}
\newcommand{\eea}{\end{eqnarray}}
\newcommand\restr[2]{{\left.\kern-\nulldelimiterspace#1\vphantom{\big|}\right|_{#2}}}

\def\eps{\epsilon}
\newcommand{\beq}{\begin{equation}} 
\newcommand{\eeq}{\end{equation}}

\def\geq{\geqslant}
\def\leq{\leqslant}

\def\eps{\epsilon}
\begin{document}
\title{Coupled minimal models revisited
}
\author{Ant\'onio Antunes$^{a,b}$, Connor Behan$^c$} 
\preprint{DESY-22-191}
\affiliation{
$^a$  Centro de F\'isica do Porto, Departamento de F\'isica e Astronomia,
Faculdade de Ci\^encias da Universidade do Porto,
Rua do Campo Alegre 687, 4169-007 Porto, Portugal\\
$^b$ Deutsches Elektronen-Synchrotron DESY, Notkestr. 85, 22607 Hamburg, Germany\\
$^c$ Mathematical Institute, University of Oxford, Andrew Wiles Building,
Radcliffe Observatory Quarter, Woodstock Road, Oxford, OX2 6GG, U.K.
}

\begin{abstract}
We study coupled unitary Virasoro minimal models in the large rank ($m \to \infty$) limit. In large $m$ perturbation theory, we find two non-trivial IR fixed points which exhibit irrational coefficients in several anomalous dimensions and the central charge. For $N>4$ copies, we show that the IR theory breaks all possible currents that would otherwise enhance the Virasoro algebra, up to spin 10. This provides strong evidence that the IR fixed points are examples of compact, unitary, irrational CFTs with the minimal amount of chiral symmetry.
We also analyze anomalous dimension matrices for a family of degenerate operators with increasing spin. These display further evidence of irrationality and begin to reveal the form of the leading quantum Regge trajectory.
\end{abstract}
\maketitle
\nopagebreak

{\bf Introduction.}
The perturbative renormalization group (RG) is a robust tool for demystifying the space of conformal field theories (CFTs) by showing how one fixed point may be reached by deforming another. In recent years \cite{os17,rs18,cmvz19,cmvz20,os20,ht20}, hundreds of new CFTs have been shown to arise from a common starting point: a tensor product of $N$ copies of the massless free scalar. In this Letter, we explore the analogous situation in which the tensor product is applied to a different type of exactly solved theory: a unitary Virasoro minimal model.

It is typically hard to make sharp statements about RG flows emanating from minimal interacting CFTs in the ultraviolet (UV). To mention a widely studied example, $N$ Ising models coupled as in
\begin{equation}
S = \sum_{i = 1}^N S^i_{\mathrm{Ising}} + g \int \mathrm{d}^d x \sum_{i < j} \epsilon^i \epsilon^j \label{coupled-ising}
\end{equation}
can be driven to a non-trivial infrared (IR) fixed point in $d = 3$. This is the critical $O(2)$ model for $N = 2$ and the hypercubic fixed point (having $\mathbb{Z}_2^N \rtimes S_N$ symmetry) for $N > 2$. Although the cubic fixed point has been studied in the $4 - d$ expansion since \cite{a73}, a crucial question about it --- whether it is preferred in nature over the $O(3)$ universality class --- could not be answered until a large scale numerical bootstrap study \cite{cllpssv20} was finally completed \cite{note1}. The problem of course is that the scaling dimension of $\epsilon^i \epsilon^j$ differs from $3$ by a finite amount which makes the flow uncontrolled.

A setup similar to \eqref{coupled-ising}, based on the $q$-state Potts model in two dimensions instead of the Ising model in $d$ dimensions, faces the same problem. The $q - 2$ expansion of
\begin{equation}
S = \sum_{i = 1}^N S^i_{q-\mathrm{Potts}} + g \int \mathrm{d}^2 x \sum_{i < j} \epsilon^i \epsilon^j \label{coupled-potts}
\end{equation}
used in \cite{djlp98} is uncontrolled because $\epsilon^i \epsilon^j$ has dimension $\Delta = \frac{8}{5} < 2$ in the most interesting case of $q = 3$ \cite{note2}. This means the maximal chiral algebra realized by \eqref{coupled-potts} in the IR along with the number of primary operators in its spectrum are both unknown.

CFTs with a finite number of primary operators are called \textit{rational}. This is because a modular invariant partition function can only be written as a finite bilinear combination of characters if the central charge and all conformal weights are rational numbers \cite{v88}. A condition weaker than rationality is \textit{compactness} defined as discreteness of the spectrum. The literature is replete with long lists
of rational CFTs \cite{klp21,mr22}. The amount of attention paid to compact irrational CFTs pales in comparison to the point where essentially all known unitary examples can be described in three lines.
\begin{enumerate}
\item The compact free boson with a generic radius. \vspace{-0.2cm}
\item Calabi-Yau sigma models with generic moduli. \vspace{-0.2cm}
\item Spinning top CFTs \cite{hkoc95} with finite fusion rules.
\end{enumerate}
All of these theories have enhanced chiral symmetry. Said another way, there are Virasoro primaries at infinitely many spins $\ell$ such that the twist $\tau \equiv \Delta - \ell$ vanishes.

For analytic bootstrap methods which apply to CFTs in $d > 2$, the presence of a twist gap
is indispensible \cite{fkps12,kz12}. Corrections to a theory's universal behaviour at large spin may be computed systematically because the contribution of a given operator to a 4pt function in the lightcone limit decays with twist. Virasoro symmetry is already enough to kill a naive application of this method in $d = 2$ but \cite{Kusuki:2018wpa,cgmp18} found a suitable improvement. Their insight was to reorganize a more modern version of the analytic bootstrap \cite{c17,lprs18} in terms of Virasoro primaries using the crossing kernel found in \cite{pt99,pt00}. This Letter aims to describe an RG flow which ends on a CFT satisfying the assumptions of the Virasoro analytic bootstrap.

This will be accomplished by regarding minimal models as distinguished points on a continuous line. In contrast to \eqref{coupled-potts} which was strongly coupled at $q = 3$ and non-unitary otherwise, our flows will become unitary and weakly coupled at the same time as we take the central charge $c \to 1$. Analytic continuation in $c$ has previously been used as a tool for interpreting numerical bootstrap results \cite{lrv12,b17}. Before this, it was implicitly used in perturbative studies of the integrable flows connecting minimal models \cite{Zamolodchikov:1987ti,Ludwig:1987gs}. Applications of this strategy to coupled systems are
scarce \cite{note3}. In the only example we are aware of, the authors of \cite{dns01} found perturbative flows
by coupling minimal models of $\mathfrak{d}_n$ type W-algebras \cite{note4}. Some evidence was found for irrationality but not for the presence or absence of enhanced chiral symmetry. The techniques developed here for the Virasoro case appear well poised for answering this question in the future.

{\bf The model.}
We start from $N$ rank $m$ unitary Virasoro minimal models with central charge $c=1-6/(m(m+1))$ each. Having in mind the large $m$ regime, where perturbation theory is well defined and where an infinite number of unitary theories accumulate, the holomorphic dimension of a primary labelled by $r, s \in \mathbb{N}$ is
\begin{equation}
h_{(r,s)} = \frac{(r-s)^2}{4} + \frac{r^2-s^2}{4m} + O(m^{-2})\,. \label{eq:largemh}
\end{equation}
To find relevant operators, we can only take pairs of the type $(r,r+1)$ and $(r,r+2)$. In the first case, we need to multiply four copies
for near marginality, but in the second case one copy suffices. This is an infinite set of deformations, but fortunately there is a finite subsector. By taking $r=1$, repeated operator product expansions (OPEs) will only produce
$(1,s)$, meaning we can truncate to products of four $(1,2)$ operators and a single $(1,3)$. The natural choice to preserve the $S_N$ symmetry permuting the copies
leads us then to the formal action
\begin{align}
	&S_{\textrm{CMM}}= \sum_{i=1}^{N} S_m^i + g_\eps \int \textrm{d}^2x N^{-\frac{1}{2}} \sum_{i=1}^{N} \phi_{(1,3)}^i \label{eq:cmm} \\
	&+ g_\sigma \int \textrm{d}^2x \binom{N}{4}^{-\frac{1}{2}} \sum_{ i<j<k<l}^N \phi_{(1,2)}^i\phi_{(1,2)}^j\phi_{(1,2)}^k\phi_{(1,2)}^l \, .\nonumber
\end{align}
The first line describes $N$
copies of the famous flow from \cite{Zamolodchikov:1987ti,Ludwig:1987gs} while the second
is more interesting.
For convenience, we will henceforth denote
these deformations by $\eps$ and $\sigma$ respectively. Subsequent analysis will use large $m$ limits of the OPE coefficients from \cite{df84,df85},
\begin{equation}
C^{(1,3)}_{(1,2)(1,2)} = \frac{\sqrt{3}}{2}, \quad C^{(1,3)}_{(1,3)(1,3)} = \frac{4}{\sqrt{3}}. \label{eq:largemc}
\end{equation}

To address the global symmetry $G$ of this model, two viewpoints are possible. If $S_m^i$ in \eqref{eq:cmm} represents only the closed subsector of $(1,s)$ operators, there is a $\mathbb{Z}_2$ symmetry sending $\phi^i_{(1, 2)} \mapsto -\phi^i_{(1, 2)}$ for all $m$ but no modular invariance. In this case, $G$ is the diagonal $\mathbb{Z}_2 \times S_N$ for $N > 4$ and $\mathbb{Z}_2^3 \rtimes S_4$ for $N = 4$.
If $S_m^i$ is instead the modular invariant theory (which only exists for integer $m$) the appropriate $\mathbb{Z}_2$ symmetry sends $\phi^i_{(1, 2)} \mapsto (-1)^m \phi^i_{(1, 2)}$ \cite{rv98} so that $G$ is hypercubic for even $m$ \cite{note5}.

{\bf Renormalization group.}
\begin{figure}
\includegraphics[scale=0.7]{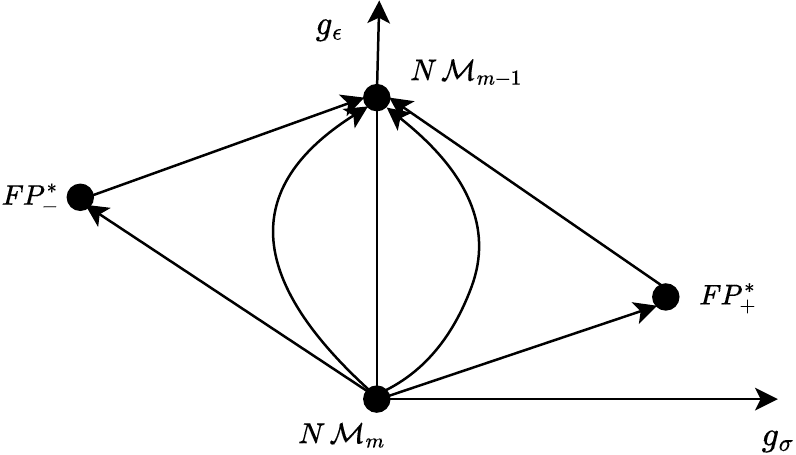}
\caption{Schematic flow diagram linking the fixed points.}
\label{fig:rg}
\end{figure}
It is straightforward to analyze the IR fixed points using conformal perturbation theory \cite{Zamolodchikov:1990bk}. For the deformation $\int \textup{d}^2x \, g^I \mathcal{O}_I$, classic one-loop results (with the summation convention) are \cite{Zamolodchikov:1986gt,Cardy:1988tj}
\begin{align}
\beta^I &= 2\tilde{g}^I - \pi C^I_{JK} g^J g^K \label{eq:general-beta} \\
\Delta c &= -2\pi^2 \mathcal{N}_{IJ} g^J \left ( 3\tilde{g}^I - \pi C^I_{KL} g^K g^L \right ) \nonumber
\end{align}
where $\tilde{g}^I \equiv (1 - h_I)g^I$ and $\mathcal{N}_{IJ} \equiv \left < \mathcal{O}_I(0) \mathcal{O}_J(\infty) \right >$. Combining \eqref{eq:largemc} and \eqref{eq:general-beta} with combinatorial gymnastics yields
 \begin{align}
	\beta_\sigma &= \frac{6}{m}g_\sigma -\frac{4\pi\sqrt{3}}{\sqrt{N}} g_\sigma g_\epsilon-6\pi \binom{N-4}{2}\binom{N}{4}^{-\frac{1}{2}} g_\sigma^2\,, \nonumber\\
	\beta_\epsilon &= \frac{4}{m}g_\epsilon - \frac{4\pi}{\sqrt{3N}} g_\epsilon^2 - \frac{2 \pi \sqrt{3}}{\sqrt{N}} g_\sigma^2
\end{align}
to leading order in $1/m$ \cite{note6}.
The beta functions have four roots.
Along with the UV fixed point $\{g_\epsilon^*=0,g_\sigma^*=0\}$ and the $N$ decoupled rank $m-1$ models $\{g_\epsilon^*=\frac{2 \sqrt{3}}{m \pi},g_\sigma^*=0\}$, there are two fully coupled fixed points $FP^*_\pm$ with
\begin{align}
	g_{\sigma\pm}^*= \pm \frac{\sqrt{(N-3)_4}}{\pi m \sqrt{2 P(N)}}\,,\, g_{\epsilon\pm}^*= \frac{\mp Q(N)+ \sqrt{3 P(N)}}{2 \pi m \sqrt{P(N)/N}}\,, \label{eq:fixed-points}
\end{align}
where $P(N)=3 N^4-53 N^3+357 N^2-1069 N+1194$ and $Q(N)=3N^2-27 N+60$. One can then also extract the IR dimensions of the deforming operators by diagonalizing the matrix $\partial \beta^I/\partial g^J$, finding the linear combinations which are dilation eigenstates in the process. The result is a rather cumbersome formula for general $N$, but
the lowest lying examples are $\Delta = 2 \pm \frac{2\sqrt{6}}{m}$ in both fixed points with $N = 4, 5$. These become $\Delta = 2 - \frac{\sqrt{6} \pm \sqrt{870}}{6m}$ for $FP^*_+$ and $\Delta = 2 + \frac{\sqrt{6} \pm \sqrt{870}}{6m}$ for $FP^*_-$ when $N = 6$.
In fact, one always gets a relevant and an irrelevant operator in the IR, so these are tricritical fixed points, as shown in figure \ref{fig:rg}.
Furthermore, for general $N$ one always finds irrational numbers multiplying $1/m$ which is a mild hint of irrationality \cite{addition}.
Indeed, if the IR CFTs were rational for every large integer $m$, with rational scaling dimensions, a natural possibility would be for the dimensions to admit an expression as a rational function of $m$. The large $m$ expansion of such a function would have rational coefficients as well. 

From the second line of \eqref{eq:general-beta},
\begin{equation}
	\Delta c_\pm^*=-\frac{2N}{m^3}\frac{\sqrt{3}Q(N) \mp 3\sqrt{P(N)}}{\sqrt{P(N)}}.
\end{equation}
These shifts are rational and the same for both fixed points in the cases $N=4,5$, but are irrational and different for $N\geq6$, showing the same type of hint.

{\bf Lifting of currents.} 
Before the interaction is turned on, the chiral symmetry of $N$ minimal models is $\mathfrak{Vir}^N$ with each factor generated by a quasi-primary stress tensor $T^i$. If the only chiral algebra surviving in the IR is a single Virasoro algebra (which is our claim), it must be the diagonal one generated by $\widehat{T} \equiv \sum_i T^i$ which we will call $\widehat{\mathfrak{Vir}}$. Any CFT with this property is guaranteed to be irrational due to the standard result that $c > 1$ implies infinitely many Virasoro primaries \cite{note7}.

We will start by proving a weaker statement --- that the IR chiral algebra is strictly smaller than $\mathfrak{Vir}^N$. The most convincing way to do this is to compute the anomalous dimension matrix for $T^i$ operators. In our case, it will come from $\sigma$ since $\epsilon$ does not couple any minimal models. One-loop conformal perturbation theory instructs us to compute $\left < T^i \sigma T^j \right >$ which vanishes by chirality while the two-loop calculation involving $\left < T^i \sigma \sigma T^j \right >$ is technically challenging.
Fortunately, we can use the alternative method of multiplet recombination \cite{Rychkov:2015naa,Giombi:2016hkj,Behan:2017emf,Behan:2017dwr}. If a short spin $\ell$ current is broken, it becomes long. For consistency with the counting of states this must happen by ``eating'' a spin $\ell-1$ divergence $V_{\ell}$:
\begin{equation}
	\bar{\partial}T_\ell=\tilde{b}(g_\sigma)V_{\ell},
\end{equation}
where $\tilde{b}(g_\sigma) = bg_\sigma + O(g_\sigma^2)$ by UV conservation. Since
\begin{align}
b \left < V_{\ell}(z_1)V_{\ell}(z_2) \right >,\quad \int \textup{d}^2z \left < \bar{\partial}T_\ell(z_1) V_{\ell}(z_2) \sigma(z) \right >
\end{align}
are both valid expressions for $g^{-1}_\sigma \left < \bar{\partial} T_\ell V_{\ell} \right >$, $b$ can be found by integrating a 3pt function \cite{note8}. As long as it is non-zero, the two-loop dimension will be given by the formula in \cite{Giombi:2016hkj} involving the UV 2pt functions of $T_\ell$ and $V_{\ell}$ \cite{note9}.

For the $N$ individual stress tensors, our ability to lift $N - 1$ degrees of freedom is strongly suggested by the fact that the operators $L^i_{-1} \sigma$ (which sum to a descendant) have the same quantum numbers as $\bar{\partial} T^i$. More precisely,
the unique divergence candidates for $T^i$ are 
\begin{equation}
	\label{eq:spin1divs}
	V^i= \sum_{(j<k<l)\neq i}(\partial \phi^i) \phi^j\phi^k\phi^l - \frac{1}{4} \partial(\phi^i \phi^j\phi^k\phi^l)\,,
\end{equation}
summing to zero, where we used the shorthand notation $\phi \equiv \phi_{(1,2)}$.
We can then compute $\langle V^i V^j \rangle$ and use the Ward identities for $T^i$ to fix $b$.
Diagonalizing this matrix, the spin 2 dilation eigenstates are $\widehat{T}$ and $T^i - T^{i + 1}$ for $i \leq N - 1$.
The anomalous dimension matrix correspondingly has a zero eigenvalue, associated to $\widehat{T}$, and the $(N-1)$-fold degenerate
\begin{equation}
	\label{eq:spin2-lift}
	\gamma[T^i - T^{i + 1}]=(g^*_\sigma \pi)^2 \frac{3}{N-1}\,
\end{equation}
describing a $\widehat{\mathfrak{Vir}}$ primary in the standard representation of $S_N$.

Moving onto the harder task, ruling out enhanced symmetry means proving that all of the infinitely many higher spin UV currents outside the $\widehat{\mathfrak{Vir}}$ identity multiplet lift in the IR. We will obtain evidence for this with a brute force check up to spin 10. The necessary computational resources can be greatly reduced by checking $\widehat{\mathfrak{Vir}}$ primaries and realizing that descendants of them will lift in the same way. Similarly, it is enough to consider $S_N$ singlets \cite{note10} which enable a compact notation. The state associated to $\widehat{T}$ in radial quantization is clearly $\sum_i L^i_{-2} \left | 0 \right >$. Suppressing indices and the vacuum, it becomes $\Sigma L_{-2}$. When encountering a multiple sum, we will implicitly subtract traces so that a product of sums includes only terms where the indices differ.
%As an example,
An example from \eqref{eq:cmm} is $\sigma = \frac{1}{4!} \binom{N}{4}^{-1/2} (\Sigma \phi)^4$. Now,
\begin{equation}
\label{eq:spin4-current}
T_4 = \Sigma L_{-4} - \frac{5}{3} \Sigma L_{-2}^2 + \frac{9}{N - 1}(\Sigma L_{-2})^2
\end{equation}
is the unique singlet primary current at spin 4. At generic $N$, the space of potential divergences is two dimensional and we can find a linear combination $V_4^\perp$ such that $\left < T_4 V^\perp_4 \sigma\right > = 0$. Its orthogonal partner
\begin{align}
V_4 &= 12(\Sigma \phi)^3(\Sigma L_{-3} \phi) - 18(\Sigma \phi)(\Sigma L_{-1} \phi)^3 \label{eq:spin4-div} \\
&+ 9(\Sigma \phi)^2(\Sigma L_{-1} \phi)(\Sigma L_{-1}^2 \phi) - 7(\Sigma \phi)^3(\Sigma L_{-1}^3 \phi) \nonumber
\end{align}
then ensures the lift of $T_4$ for general $N$ \cite{note11} with
\begin{equation}
	\gamma[T_4] = (g^*_\sigma \pi)^2 \frac{5N + 22}{2N(N - 1)}\,. \label{eq:spin4-lift}
\end{equation}

Table \ref{tab:currents-food} counts $T_\ell, V_{\ell}$ operators for increasing spin. As with the counting in \cite{b18}, the matrix $\left < T_\ell^I V_\ell^J \sigma \right >$ quickly becomes much wider than it is tall \cite{note12}. This makes it highly believable that the rows will be linearly independent. Code which performs the explicit check is attached to this Letter's arXiv submission \cite{note13}. At $\ell = 10$, we have run it for several values of $N \geq 5$ (which takes about one CPU day) and found that everything lifts. For $\ell \leq 8$, we have additionally done a symbolic check which establishes this result for all values of $N$ which are large enough for the numbers in Table \ref{tab:currents-food} to stabilize.
\begin{table}[h] 
\centering
\begin{tabular}{c|cccc}
\backslashbox{$\ell$}{$N$} & 4 & 5 & 6 & 7 \\
\hline
4 & (1, 1) & (1, 2) & (1, 2) & (1, 2) \\
6 & (2, 2) & (2, 5) & (2, 6) & (2, 6) \\
8 & (4, 7) & (4, 17) & (4, 22) & (4, 23) \\
10 & (5, 18) & (7, 50) & (7, 69) & (7, 75)
\end{tabular}
\caption{Ordered pairs giving the number of primary singlet currents $T_\ell$ and then the number of potential divergences for them $V_{\ell}$ built out of $\phi_{(1, 2)}$.}
\label{tab:currents-food}
\end{table}
The case of exactly four copies is a different story. The $2 \times 2$ matrix for $\ell = 6$ has zero determinant which means a current at this spin is conserved to two loops \cite{note14}. Signs of enhanced symmetry therefore appear if and only if $N = 4$.

{\bf Double twist operators.}
\begin{figure}
\includegraphics[scale=0.6]{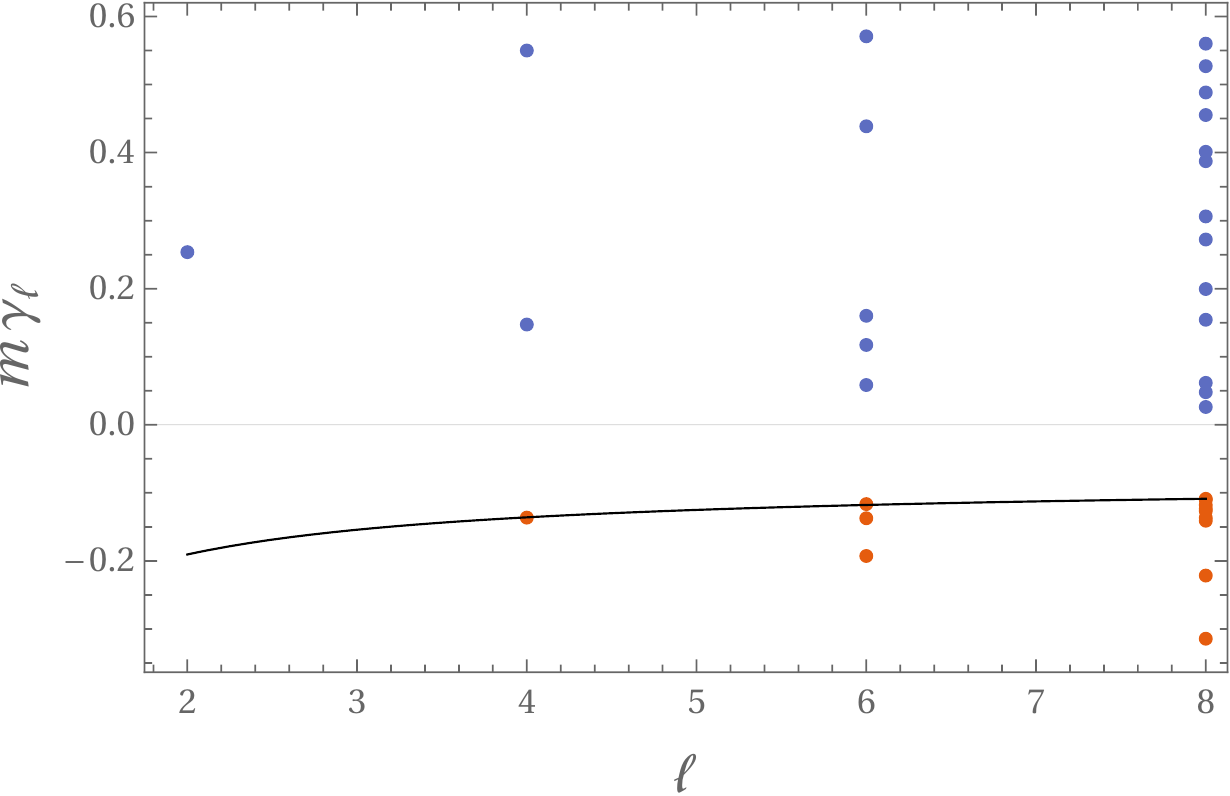}
\caption{Anomalous dimensions of types 1 (blue) and 3 (orange) at $N = 14$ showing one possible interpolation.}
\label{fig:regge}
\end{figure}
Given our handle on the multiplet of conserved currents, analyticity in spin \cite{c17} makes it tempting to look for trends within operator families. Consider $\Phi \equiv \Sigma \phi$ with $\widehat{\mathfrak{Vir}}$ primaries
\begin{align}
\mathcal{O}_0 &= (\Sigma \phi)^2, \label{eq:dt-spin02} \\
\mathcal{O}_2 &= (\Sigma \phi)(\Sigma \bar{L}_{-1}^2 \phi) - 3(\Sigma \bar{L}_{-1} \phi)^2 - \frac{3}{N - 2}(\Sigma \bar{L}_{-2})(\Sigma \phi)^2, \nonumber
\end{align}
of increasing spin ($\bar{T}$ weight) in $\Phi \times \Phi$. For $m \to \infty$, these clearly have unit twist. Another regime which makes the twist of \eqref{eq:dt-spin02} well understood is $N \to \infty$. Even though this sends \eqref{eq:spin4-lift} to zero, it also causes all higher Virasoro generators to decouple leaving the $\mathfrak{sl}(2)$ result $\tau \to 1 + 2\gamma_\Phi \approx 1 - \frac{1}{2mN}$ as $\ell \to \infty$ \cite{fkps12,kz12}. This limit should be reached monotonically due to Nachtmann's theorem \cite{n73}. To investigate smaller values of $m$ and $N$, we have computed anomalous dimensions up to spin 8 leading to matrices of size $1$, $1$, $4$, $11$ and $29$. Employing
\begin{equation}
(\Gamma)_I^{\;\;J} = -\pi \left ( g_\sigma C_{IK\sigma} + g_\eps C_{IK\eps} \right ) (\mathcal{N}^{-1})^{KJ} \label{eq:adm}
\end{equation}
to accommodate operators which are not orthonormal \cite{note15}, two checks become apparent. First, the $\eps$ term is provably a multiple of the identity as required for a decoupled flow. Second, eigenvalues of the $\sigma$ term \cite{note16} (and hence of the sum) appear to be solvable by radicals if and only if $N = 4$. Some $N = 4$ eigenvalues are degenerate thus demonstrating the effect of the spin 6 current \cite{note17}.

More extensive numerical experiments show that 45 of these 46 eigenvalues fall into the following three types.
\begin{enumerate}
\item Positive numbers decaying as $N^{-1}$ for large $N$. \vspace{-0.2cm}
\item Negative numbers with a finite large $N$ limit. \vspace{-0.2cm}
\item Negative numbers asymptotic to $-\frac{1}{2mN}$ \cite{note18}.
\end{enumerate}
Type 3 is remarkable because it shows that $\tau = 1 + 2\gamma_\Phi$ can be achieved at large $N$ without the spin being large. Let us therefore keep track of where type 3 eigenvalues move as we lower $N$ to $14$. Assuming a small shift in the central charge, this is the smallest integer allowing the Virasoro identity block in $\Phi \times \Phi$ to have a discrete contribution in the crossed channel. These contributions were termed \textit{quantum Regge trajectories} in \cite{cgmp18} and shown to reproduce ordinary Regge trajectories from mean field theory as $c \to \infty$. At large spin, their twists are bounded above by $\frac{c - 1}{12}$ and satisfy a version of Nachtmann's theorem. If we assume the unique type 3 eigenvalue at spin 4 lies on such a trajectory, its fate for higher spins is tightly constrained by monotonicity. At spin 6, there is only one type 3 eigenvalue which can give it a larger twist. At spin 8, there are two but the distance between them is very small. The picture that emerges is Figure \ref{fig:regge}.

These perturbative results can no longer be trusted when $\ell$ becomes large enough to compete with $m$. The behaviour which takes over is found by relating $\tau$ to $c$ and $\Delta_\Phi$ with the Virasoro analytic bootstrap \cite{Kusuki:2018wpa,cgmp18} and then perturbing the latter quantities. The result
% The shift in c is not needed at this order
\begin{equation}
\tau = \frac{\sqrt{N - 1} - \sqrt{N - 7}}{3 / \sqrt{N - 7}} + \frac{2\pi g^*_\eps}{\sqrt{N / 3}} \left ( \sqrt{\frac{N - 1}{N - 7}} - 2 \right ) \label{eq:vir-boot}
\end{equation}
holds for $\sqrt{\ell} \gamma^* \gg 1$ or $\sqrt{\ell} \gg m \gg 1$. Here, $\gamma^*$ is the twist of the most weakly broken current which we expect to be $T_4$. It would be interesting to find a non-perturbative estimate for this Regge trajectory in between the regimes of \eqref{eq:vir-boot} and Figure \ref{fig:regge}.

{\bf Discussion.}
Due to the recombination analysis in this Letter, Occam's razor favours the following scenario. \textit{The fixed points \eqref{eq:fixed-points} with $N > 4$ have only Virasoro symmetry and are therefore irrational.} If this were false, the first $S_N$ singlet disproving it would need to have a spin of at least 12. This conclusion should also apply to any extension of \eqref{eq:cmm} which explicitly breaks $S_N$ \cite{note19}. In particular, there are simple interactions preserving $\mathbb{Z}_N$ which is the symmetry
of a stack of layers with periodic boundary conditions.
Following \cite{llm97}, the large $N$ limit in such cases could give a window onto three dimensional physics.

All models just discussed can be defined for the continuum or the lattice. Hamiltonian truncation and Monte Carlo techniques are therefore both available for determining the precise extent of the conformal window \cite{note20}. 

Finally, there is much that can be said about analogues of \eqref{eq:cmm} which couple minimal models of a W-algebra. The setup examined in \cite{dns01} uses $W[\mathfrak{d}_n]$ which is part of a family $W[\mathfrak{g}]$ labelled by a simply laced Lie algebra. If one includes the A-series as well, the space of $S_N$ preserving flows becomes richer but not infinitely so. In particular, the requirement that operators become marginal as $c \to \mathrm{rank}(\mathfrak{g})$ allows $\mathfrak{g}$ to be no larger than $\mathfrak{a}_8$. A detailed study of the various possibilities will appear in future work \cite{abfuture}.

\vspace{10pt}

\begin{acknowledgments}
We are grateful to the Simons Foundation where this work was initiated and to the Galileo Galilei Institute where part of it was presented. We also thank C. Beem, S. Collier, M. Costa, A. Kaviraj, S. Rychkov, V. Schomerus, B. van Rees, X. Yin and B. Zan for useful discussions.
AA was partially supported by a Simons Collaboration on the Non-Perturbative Bootstrap scholarship in the University of Porto and received funding from the German Research Foundation DFG under Germany’s Excellence Strategy – EXC 2121 Quantum Universe – 390833306. CB received funding from the European Research Council (ERC) under the European Union's Horizon 2020 research and innovation programme (grants 682608 and 787185).
\end{acknowledgments}


\begin{thebibliography}{99}
\bibitem{note1} See also \cite{b21} for a recent perturbative approach which takes numerical bootstrap data as input.
\bibitem{note2}{A further problem, reviewed in \cite{grz18}, is that some of the 3pt functions at $g = 0$ affecting higher loop corrections are not yet known for general $q$.}
\bibitem{note3} A related case in \cite{bdlv21} couples minimal models not to each other but to the boundary mode of a free scalar living in a higher dimensional bulk.
\bibitem{note4} The motivation for considering this W-algebra is that it leads to a quadratic interaction resembling \eqref{coupled-ising} and \eqref{coupled-potts} in contrast to the interactions we will discuss which are quartic. The term $\epsilon^i \epsilon^j$ is natural in the study of disordered systems where it represents the effect of integrating out a random bond strength with a Gaussian distribution. If observables in such a model can be understood as analytic functions of $N$, disorder averages follow by taking $N \to 0$ \cite{ks16}.
\bibitem{note5} Analyticity in $m$ is also more straightforward within the $(1, s)$ subsector than it is within the full theory. For an operator with Kac indices $r, s \in \mathbb{N}$, the condition $r, s \ll m$ is not only needed for the IR correlators involving this operator to be accessible with perturbation theory. If $r > 1$ and $s > 1$ then even before we flow, there is an upper limit to the number of times $\sigma$ and $\epsilon$ can be inserted before the UV correlator develops a discontinuity as $m$ approaches an integer. The results in \cite{r18} for 4pt functions make it possible to determine how many insertions are safe by iteratively applying the OPE.
\bibitem{note6} The strictly relevant operator $\sum_{i<j} \phi_{(1,2)}^i \phi_{(1,2)}^j$ is also generated but can be tuned to zero with impunity to this order in conformal perturbation theory.
\bibitem{addition} The coefficients once again approach rational numbers in the large $N$ limit. The weakly relevant and weakly relevant eigenstates have $\Delta = 2 + \frac{1 \pm 5}{m} + O(N^{-1})$ for $FP^*_+$ and $\Delta = 2 - \frac{1 \pm 5}{m} + O(N^{-1})$ for $FP^*_-$.
\bibitem{note7} From the point of view of our UV starting point, $\mathfrak{Vir}^N$ characters contain an infinite number of diagonal Virasoro characters when decomposed.
\bibitem{note8} This consists of integrating a pure contact term. When evaluated correctly, the 3pt function has the Polyakov form which makes it straightforward to use the identity $\bar{\partial}(z_1 - z)^{-1} = 2\pi\delta(z_1 - z, \bar{z}_1 - \bar{z})$.
\bibitem{note10} A conserved higher spin current in another representation will produce a conserved singlet after normal ordered products are taken
\bibitem{note11} For $N=4$, consistently with Table \ref{tab:currents-food}, $V^\perp_4$ does not exist since the terms $(\Sigma L_{-3})(\Sigma \phi)^4$ and $(\Sigma L_{-2})(\Sigma \phi)^3(\Sigma L_{-1} \phi)$ are quintic. The relevant space is then spanned by \eqref{eq:spin4-div} to begin with.
\bibitem{note12} The particular divergence associated to each current is not uniquely fixed when this matrix is not square. These need to be determined if one wants precise values for anomalous dimensions. Here, we are only concerned with showing that they are non-zero.
\bibitem{note13} Operators used in the check are constructed by applying $\widehat{L}_1, \widehat{L}_2$ to the most general ansatz. Each term therein is a product of up to $N$ sums and the sums are populated with negative modes and $\phi$ in all possible ways that yield $h = \ell$. Inside each sum (sorted with $\phi < L_{-1} < L_{-2} < \dots$), there are restrictions coming from null states. No sum can contain $\phi$ twice and any sum with (without) $\phi$ is forbidden from containing $L_{-2}$ ($L_{-1}$).
\bibitem{note14} It is possible that every other higher spin current admits an expression as a normal ordered product involving this strong generator of spin 6. If this turns out to be the case, the fixed points \eqref{eq:fixed-points} for $N = 4$ will have chiral symmetry given by $\mathcal{W}(2, 6)$ \cite{bfknrv91}. Minimal model central charges of this algebra are known and do not accumulate at $4$ \cite{efhhnv92}.
\bibitem{note15} If one instead chooses operators which are orthonormal in the UV, this property will be preserved by dilation eigenstates in the IR since the OPE coefficients in \eqref{eq:adm} are symmetric.
\bibitem{note16} See \cite{llm97} which shows that the pure $\sigma$ deformation (which does not reach a fixed point) is integrable for $N = 4$.
\bibitem{note17} This phenomenon has even been used to discover new symmetries in \cite{ct18}.
\bibitem{note18} In this section, we are considering the $FP^*_+$ fixed point from \eqref{eq:fixed-points} rather than $FP^*_-$ which is dominated by $g^*_\eps$ at large $N$.
\bibitem{note19} Among the fixed points of these more general models, we expect that many will simply recycle results for smaller values of $N$ and need to be filtered out \cite{os17}.
\bibitem{note20} A naive guess is that it might persist all the way down to $m=4$, which would be compatible with the topology of the large $m$ phase diagram, and is indeed what happens for the pure $\epsilon$ deformation (see also \cite{bdlv21}).
\end{thebibliography}
\end{document}